\documentclass[aps,floatfix,twocolumn,a4paper,showpacs, nofootinbib,superscriptaddress,10pt]{revtex4}
\usepackage{graphicx,float}\usepackage{graphicx,float}
\usepackage[all]{xy}
\usepackage{amsmath,upgreek}
\usepackage{amssymb}
\usepackage{color}
\usepackage{epsfig,bm}		
\usepackage{graphicx,epstopdf}
\usepackage{subfigure}
\usepackage{pdfpages,slashed}
\usepackage[colorlinks,hyperindex]{hyperref}

\definecolor{green1}{RGB}{0,128,0}
\hypersetup{backref=true,pagebackref=true}
\hypersetup{%
  colorlinks = true,
  linkcolor  = blue,
  citecolor = cyan,
}
\usepackage{bookmark,textgreek}
\usepackage{hyperref,color,xcolor}
\hypersetup{hidelinks,hyperindex=true,colorlinks=true,breaklinks=true,urlcolor= blue}
\hypersetup{%
  colorlinks = true,
  linkcolor  = blue
}\usepackage{amssymb}
\newsavebox{\foobox}

\usepackage{graphicx,float,tikz}
\usepackage[all]{xy}
\newcommand\ringring[1]{%
  {
   \mathop{\kern0pt #1}\limits^{
     \vbox to-1.85ex{
       \kern-2ex 
       \hbox to 0pt{\hss\normalfont\kern.1em \r{}\kern-.45em \r{}\hss}%
       \vss 
     }
   }
  }
}

\newcommand{\bpartial}{\mathop{\partial\kern -4pt\raisebox{.8pt}{$|$}}}

\newcommand{\bes}{\begin{subequations}}
\newcommand{\ees}{\end{subequations}}
\def\beq{\begin{eqnarray}}

\def\eeq{\end{eqnarray}}
\def\be{\begin{equation}}
\def\ee{\end{equation}}

\begin{document}

\title{Total hadronic and photonic cross sections and the nuclear configurational entropy concept}
\author{G. Karapetyan}
\email{gayane.karapetyan@ufabc.edu.br}
\affiliation{Federal University of ABC, Center of Natural Sciences, Santo Andr\'e, 09580-210, Brazil}
\affiliation{Federal University of ABC, Center of Mathematics, Santo Andr\'e, 09580-210, Brazil}

\begin{abstract}
In the framework of eikonalized mini-jet model,  contributions from gluon recombination effects are studied for the free proton in $pp$ and $\bar p p$ reactions.
The nuclear configurational entropy is then used to calculate the main parameters to estimate the total proton cross sections for the recent data LHC, also corroborating to
previous developments. Photoproduction cross sections with vector meson dominance and saturation effects for both total hadronic and photonic cross sections at a high energy regime are investigated. Extreme points of the configurational entropy provide the experimentally obtained cross sections within an optimal  root-mean-square deviation.

\end{abstract}
\maketitle
\section{Introduction}

The interest towards a study of strong interaction in the Color Glass Condensate (CGC) setup has been directed to how the configurational entropy (CE) approach can indicate directions for experiments and various methods in QCD \cite{Ferreira:2019inu,Bernardini:2018uuy,daRocha:2021imz,Ferreira:2020iry,Karapetyan:2018yhm}.
Answering this question requires knowledge of the information of the total inelastic cross section that could lead to collective nuclear excitations from the point of view of the nuclear configurational entropy and its impact on the fundamental reaction mechanisms
\cite{Karapetyan:plb,Karapetyan:2020epl,Karapetyan:2021epjp}.
Analytical and numerical studies have been already devoted to the deep investigation of such techniques concerning various aspects of interaction, including  the production of mesons (scalar, vector and tensorial) resonances \cite{Braga:2018fyc,Bernardini:2016hvx,Barbosa-Cendejas:2018mng,Ferreira:2019nkz,daRocha:2021ntm,Karapetyan:2021ufz}, glueballs \cite{Bernardini:2016qit,MarinhoRodrigues:2020yzh}, charmonium and bottomonium  \cite{Braga:2017fsb}, the quark-gluon plasma \cite{daSilva:2017jay}, baryonic fields  \cite{Colangelo:2018mrt}, and also using the AdS/QCD model approach \cite{Ma:2018wtw}.
The configurational-entropic origin of the above mentioned studies can be found in  Refs.  \cite{Gleiser:2012tu, Gleiser:2011di, Gleiser:2013mga, Gleiser:2018kbq,Sowinski:2015cfa,Gleiser:2014ipa,Gleiser:2015rwa}.
Extreme  points of the CE driving dominant states of physical systems have been also studied and analyzed in Refs. \cite{Bernardini:2019stn,Correa:2015vka,Bazeia:2021stz,Casadio:2016aum,Fernandes-Silva:2019fez,Braga:2016wzx,Braga:2020myi,Alves:2017ljt,Alves:2014ksa,Correa:2015lla}, also for AdS black holes and their quantum portrait as Bose-Einstein graviton condensates, emulating magnetic structures \cite{Bazeia:2018uyg}.
These critical points manifest the stability and the
predominance of the system states, which also give the vision about the mechanism of interaction for different nuclear systems using, for instance, the CGC for parton saturation effects \cite{Karapetyan:2018yhm,Karapetyan:2019epl,Ma:2018wtw,Karapetyan:plb,
Karapetyan:2017edu,Karapetyan:2016fai}.
Such points  have been successfully investigated for the calculation of the inelastic hadron cross sections. The probability of production of different reaction channels can be suggested as spatially localized configurations at high energy regimes \cite{Karapetyan:2018yhm,Karapetyan:plb,Karapetyan:2017edu,Karapetyan:2016fai}.

As it is well known, the total hadronic cross section rises with energy, which can be explained by QCD. There is a close relationship between the total hadronic cross section at the energy determined concerning the center-of-mass energy ($\sqrt s$) and the production of
jets with the principal role of the interacted partons, comprising the so-called
minijet model \cite{Giannini}.
According to this model, the total hadronic
cross sections at high energy can be represented as a sum of  two terms: a nonperturbative energy-independent contribution, $\upsigma_0$, and a semi-hard perturbative QCD (pQCD) contribution,
$\upsigma_{pQCD}$, at low transverse momenta $p_{T_{min}}$.
A perturbative component at a high energy regime is characterized by
small fractional momentum (small Bjorken-$x$) of gluons. All calculations based on the mini-jet model confirm the rise of the hadron cross section with energy within the pQCD.
The eikonal representation of the mini-jet model involves a comprehensive description of the dynamics of partons  at high energies.
Numerical experiments results of a deep inelastic scattering from HERA in a wide range of four-momentum transfer to the proton, $Q^2$, at small $x$, confirmed the theoretical calculations of a rapid gluon density rise with a falling of $x$ value as a result of nonlinear effects in gluon evolution equations. Such a growth at $x \lesssim 10^{-4}$ can be explained as gluons domination at this range, with the achievement of saturation afterward.

There are two main cases concerning the momentum
transfer, $\emph{\textbf{k}}_{\perp}$, and the
transverse size of a gluon during $g \rightarrow gg$ interactions: on the one hand, at large momentum transfer one can expect a large number of small size gluons per unit of rapidity.
Another case is that small momentum transfer gluons, which are originated during the interaction, overlap themselves in the transverse zone via the fusion process. The fusion case has a low probability for values $\emph{\textbf{k}}^2_{\perp}>Q^2_{s}$, where $Q_{s}$ is a
saturation scale. And the probability of the fusion rise at
momentum transfer $\emph{\textbf{k}}^2_{\perp}<Q^2_{s}$
as the gluon density is large and rises with small $x$.
Within a CGC, a system is limited by the maximum phase-space parton density. This condition is described by the hadron wave function and the energy-dependent momentum
scale, $Q_{s}(x)$.

In this paper the nuclear CE concept is used to study the influence of the gluon recombination during high-energy photonic and hadronic interactions, namely, to reconstruct
experimentally obtained total reaction cross sections.
In Section II we describe the eikonalized mini-jet model for hadronic and photonic cross sections as well as present data analysis for the inelastic interaction within the Glauber formalism.
In Section III we compare the results of our calculations using the CE approach and give the basic onset for the probability of the production cross section in photon and hadron collisions from the point of view on the nuclear system configuration.
Section IV is devoted to the conclusion.

\section{Eikonalized mini-jet model}

The main idea of the hadronic cross section within the QCD is a
unitarity constraint formalism. In the so-called eikonal formulation, the mini-jet cross sections undergo the unitarization procedure after
sequential multiple scattering process.
Here the total, elastic, and inelastic
$pp (\bar p)$ cross sections can be respectively represented as
\begin{eqnarray}
\upsigma^{pp(\bar p)}_{tot} (s)&=& 2 \int d^2 \vec{b}\,
\{1-e^{-\Im\,\chi (b,s)} \cos [\Re\,\chi (b,s)]\}\,, \\
\upsigma^{pp(\bar p)}_{el} (s)&=& \int d^2 \vec{b} \,|1-e^{i\,\chi
(b,s)}|^2\,, \\
\upsigma^{pp(\bar p)}_{inel} (s)&=& \int d^2 \vec{b}
\,[1-e^{-2\,\Im\,\chi (b,s)}].
\label{sigpptot}
\end{eqnarray}
where $b$ is the impact parameter of the collision;
$\chi (b,s)$ is the eikonal function, which depends on the
energy and the transverse momentum of the nuclear matter distribution and can be determined as
$\chi (b,s) = {\Re}\, [\chi (b,s)] + i\,\Im\, [\chi (b,s)]$.
In the case of for $pp(\bar p)$ reactions, the real part of eikonal function,
${\Re}\,\chi (b,s)=0$ is suggested to be equal to zero as in the
ratio of the real to the imaginary part of the forward elastic
amplitude the real part of $\chi (b,s)$ consist of $4\%$.
The interaction of the partons obey a Poisson distribution and the average number of the inelastic collisions can be represented as a sum of soft and hard as $n(b,s) \equiv 2\,{\Im}\,\chi (b,s) =
n_{soft}(b,s) + n_{hard}(b,s)$ concerning such parameters as
$b$ and $s$:
\beq
n(b,s) &=& W(b,\upmu_{soft})\,\upsigma^{soft}(s)\nonumber \\
&&+\sum_{k,l}\,W(b,\upmu_{hard})\,\upsigma^{hard}_{kl}(s).
\label{numbcoll}
\eeq
In Eq. \eqref{numbcoll} $W(b,\upmu_{soft})$ and $W(b,\upmu_{hard})$ are the
effective overlap functions of the nucleons, which contain  information about the nucleon form factor and usually are normalized according to $\int{W(b,\upmu)\,d^{2}\vec{b}}=1$;
$\upsigma^{soft}(s)$ and $\upsigma^{hard}_{kl}(s)$ are the excitation functions of the reactions, measured in milibarns (${\rm mb}$).
While one can neglect the influence of the hard term of the eikonal function in the low energy regime, the soft contribution in the case of $pp$ and $p\bar p$ scattering at the given energy proton in
the laboratory system, $E_{lab}$, can be represented as the following:
\begin{eqnarray}
W(b,\upmu_{soft})&=& \frac{\upmu^2_{soft}}{96\pi}(\upmu_{soft}
b)^{3}K_{3}(\upmu_{soft} b),  \\
\upsigma^{pp}_{soft}(E_{lab})&=&47 + \frac{46}{E^{1.391}_{lab}},  \\
\upsigma^{p\bar{p}}_{soft}(E_{lab})&=&47 +
\frac{129}{E^{0.661}_{lab}} + \frac{357}{E^{2.72}_{lab}},
\label{W.de.b}
\end{eqnarray}
with a fitted parameter $\upmu_{soft}$, whereas $K_{3}$ is the
modified Bessel function.
Within the mini-jet leading order (LO) pQCD model, when the partons are produced back to back in the transverse plane, the differential production cross section can be given by:
\begin{widetext}
\begin{equation}
\frac{d\upsigma^{mj}_{kl}}{dy}(s)=\kappa \int dp_T^2 \,dy_2
\sum_{{i,j}}x_1\, f_{i/h_1}(x_1,Q^2) \,x_2\,f_{j/h_2}(x_2,Q^2)\,
\frac{1}{1+\delta_{kl}}
\,\left[\delta_{fk}\,\frac{d\hat{\upsigma}^{ij\rightarrow
kl}}{d\hat{t}}(\hat{t},\hat{u})+\delta_{fl}\,\frac{d\hat{\upsigma}^{ij\rightarrow
kl}}{d\hat{t}}(\hat{u},\hat{t})\right]\,
\label{eq:mj.cs}
\end{equation}
\end{widetext}
where the constant $\kappa$, $h_1$ and $h_2$ are the interacting hadrons
and by $d\hat{\upsigma}^{ij\rightarrow kl}/d\hat{t}$ one labeled the subprocess cross sections ($gg\rightarrow gg$, $gq(\bar q)\rightarrow gq(\bar q)$
and $gg\rightarrow q\bar q$ ($q \equiv u, d, s$)).
In such case, the hard term of the eikonal functions have the following  form: $W(b,\upmu_{gg})$, $W(b,\upmu_{gq} \equiv \sqrt{\upmu_{ qq}\upmu_{gg}})$ and $W(b,\upmu_{qq})$ with free mass parameters
$\upmu_{qq}$ and $\upmu_{gg}$, which determine the effective area
occupied by quarks and gluons.
For each final state of the partons, $k$ and $l$ corresponds the parameter of rapidity, $y\,(\equiv y_1)$ and $y_2$, while the transverse momentum of a parton is denoted by $p_T$ $\geq p_{T_{min}}$, where $T_{min}$ is the smallest transverse momentum of a parton with the corresponding density $f_{i,j/h_{1,2}}(x_{1,2},Q^2)$.
Moreover, for two colliding partons, $i$ and $j$, the fractional momenta are $x_{1,2} =
p_T/\sqrt{s}\,(e^{\pm y} + e^{\pm y_2})$ and the ratio
$1/(1 + \delta_{kl})$ determines a statistical weight of the particles in
the final state.

Within the eikonal limit, which implies the straight-line trajectories of
colliding nucleons, one can use the Glauber multiple collision approximation in order to estimate the inelastic proton-nucleus cross section, $\upsigma_{inel}^{pA}(s)$. Such cross section is immediately
derived from the corresponding inelastic
nucleon-nucleon ($NN$) cross section, $\upsigma_{inel}^{NN}(s)$ as:
\begin{equation}
\upsigma_{inel}^{pA}(s)= \int d^2\vec{b} \left[1 -
e^{-\upsigma_{inel}^{NN}(s)\,T_A(b)}\right],
\label{siginel.pA.final}
\end{equation}
where  $T_A(b) \equiv \int dz
\rho_A(b,z)$ is the thickness function with $\int d^2\vec{b}\,T_A(b)=A$,
which determines the number
of nucleons in the nucleus $A$ per area concerning the $z$ axis.
Then, the density of the nucleus $A$ with a radius $R_A$ can be represented as:
\begin{equation}
\rho_A(b,z)= \rho_{0}\,\{1+exp\,[(r-R_{A})/a_{0}]\}^{-1}\,
\label{woodsaxon}
\end{equation}
with $r=\sqrt{b^{2}+z^{2}}$, $R_{A}= 1.19 \,A^{1/3}-1.61
\,A^{-1/3}\,(fm)$. In Eq. \eqref{woodsaxon} $\rho_{0}$ is a density
in the center of the nucleus $A$ and $a_0$ is parameter of the diffusion in the Woods-Saxon potential, which assumed as a free parameter.

The mini-jet model is used also for the calculation of $\gamma{p}$ and $\gamma\gamma$ cross sections. One can use
$pp$ forward scattering amplitude through the vector meson dominance (VMD) to derive the cross section for  reactions as well as the additive quark model, in which one uses a
probability ($P_{had}^{\gamma{p}(\gamma)}$) that the photon
interacts as a hadron.
Thus, if we suggest that the interaction of high energy photon as a hadron, consisting of two quarks
($\upsigma^{s,h}\mapsto \frac{2}{3}\upsigma^{s,h}$ and
$\upmu_{s,h}\mapsto \sqrt{\frac{3}{2}\upmu_{s,h}}$ in both, soft and hard components of Eq. (\ref{numbcoll})),  one can derive the simple equation for $\gamma{p}$ cross section in the following form:
\begin{equation}
\upsigma^{\gamma{p}}_{tot} (s)\!=\! 2 P_{had}^{\gamma{p}} \int d^2 \vec{b}
\{1-e^{-{\Im}\chi^{\gamma{p}} (b,s)}  \cos [{\Re}\chi^{\gamma{p}} (b,s)]\}.
\label{siggammaptot}
\end{equation}
And in the case of $\gamma\gamma$ cross section (($\upsigma^{s,h}\rightarrow \frac{4}{9}\upsigma^{s,h}$
and $\upmu_{s,h}\rightarrow \frac{3}{2}\upmu_{s,h}$), one has correspondingly:
\begin{equation}
\upsigma^{\gamma\gamma}_{tot} (s)\!=\! 2P_{had}^{\gamma\gamma} \int d^2 \vec{b}
\{1-e^{-\Im\chi^{\gamma\gamma} (b,s)} \cos [{\Re}\chi^{\gamma\gamma} (b,s)]\}
\label{siggammagammatot}
\end{equation}
where $P_{had}^{\gamma\gamma}$ is a free parameter.

\section{The model parametrization and configurational entropy}

During the analysis, for instance, of the latest LHC data concerning reaction cross section for $pp$ and $p\bar p$ scattering as well as for $\gamma$-proton, and $\gamma\gamma$ reaction, one uses the fitted parameters for the soft term of the model at low energies, and the fixed parameters for the hard term at higher energies.
Then using Eqs. (\ref{siginel.pA.final}), (\ref{siggammaptot}), and (\ref{siggammagammatot}), the inelastic $p$-Air, $\gamma{p}$, and $\gamma\gamma$  cross sections for $pp$ and $p\bar p$ reactions, which is related from the energy dependence of the total cross sections for the $pp$ and $p\bar p$ reactions (\ref{sigpptot}). The fixed hard parameters for nonlinear evolution (EHKQS) were studied in Ref. \cite{Giannini} at
\beq
&&p_{T_{min}}^{2} = 1.51\, {\rm GeV}^{2}, \qquad \upmu_{gg}=2.00\, {\rm GeV},\nonumber\\ &&\upmu_{q\bar{q}} = 0.70\, {\rm GeV},\label{papar}\eeq whereas the parameter $\upmu^{2}_{soft}=0.7 \,{\rm GeV}^{2}$ (Eq. (\ref{W.de.b})) was used \cite{Giannini}. The relevant experimental data shown in the next figures considers only the most quoted ones in the literature
and can be found in the references \cite{Block:2006hy,Antchev:2013iaa,Collaboration:2012wt,pdg,Abelev:2012sea,Aielli:2009ca}.

The analysis in these Refs. yields the increment of cross sections at high energies induced by the increasing of gluon densities, at small $x$. The
nonlinear evolution leads to a small
softening of these cross section due gluon fusion
$gg \to g$ processes. Above $\sqrt{s}\sim 6$ GeV,
 the cross section in Ref. \cite{Giannini} was fitted by Froissart-like parametrizations,
\beq
\upsigma(s) &=& (30.5 \pm 0.8) + \alpha_1 s^{\beta_1}- \alpha_2 s^{\beta_2}\nonumber\\&&+ (0.2014 \pm 0.0035) \log^2(s),\eeq
 where $\alpha_1=42.53 \pm 1.35$ mb, $\alpha_2=33.34 \pm 1.04$ mb, $\beta_1=-0.458 \pm 0.017$, and $\beta_2=-0.545 \pm 0.007$.

In our analysis, the mini-jet model of QCD can be then employed for the photoproduction cross sections with the inclusion of the vector meson dominance and the saturation effects for total hadronic and photonic cross sections at a high energy regime.
The decomposition of the Fourier function into the number of weighted component modes, associated with hadronic cross sections was introduced in Ref. \cite{Bernardini:2016hvx}.
The  total inelastic cross section can be transformed through the energy-weighted correlation Fourier transform  of the cross section, considered as a spatially localized function,
\begin{equation}
\label{34}
\upsigma(k)=\frac{1}{\sqrt{2\pi}}\! \int_{\mathbb{R}}\upsigma(x)e^{ikx} d x.
\end{equation}
It is worth emphasizing that the cross section $\sigma$ is
a function of the parameters $p_{T_{min}}$, $\upmu_{gg}$, and $\upmu_{q\bar{q}}$. The important role of nuclear CE-based techniques is to try to derive from the theoretical point of view the values assumed by the parameters listed in Eq. (\ref{papar}), obtained by experiments, as the ones which can extremize the nuclear CE of the physical system.

\begin{figure}[h]
	\centering
	\includegraphics[width=5.8cm]{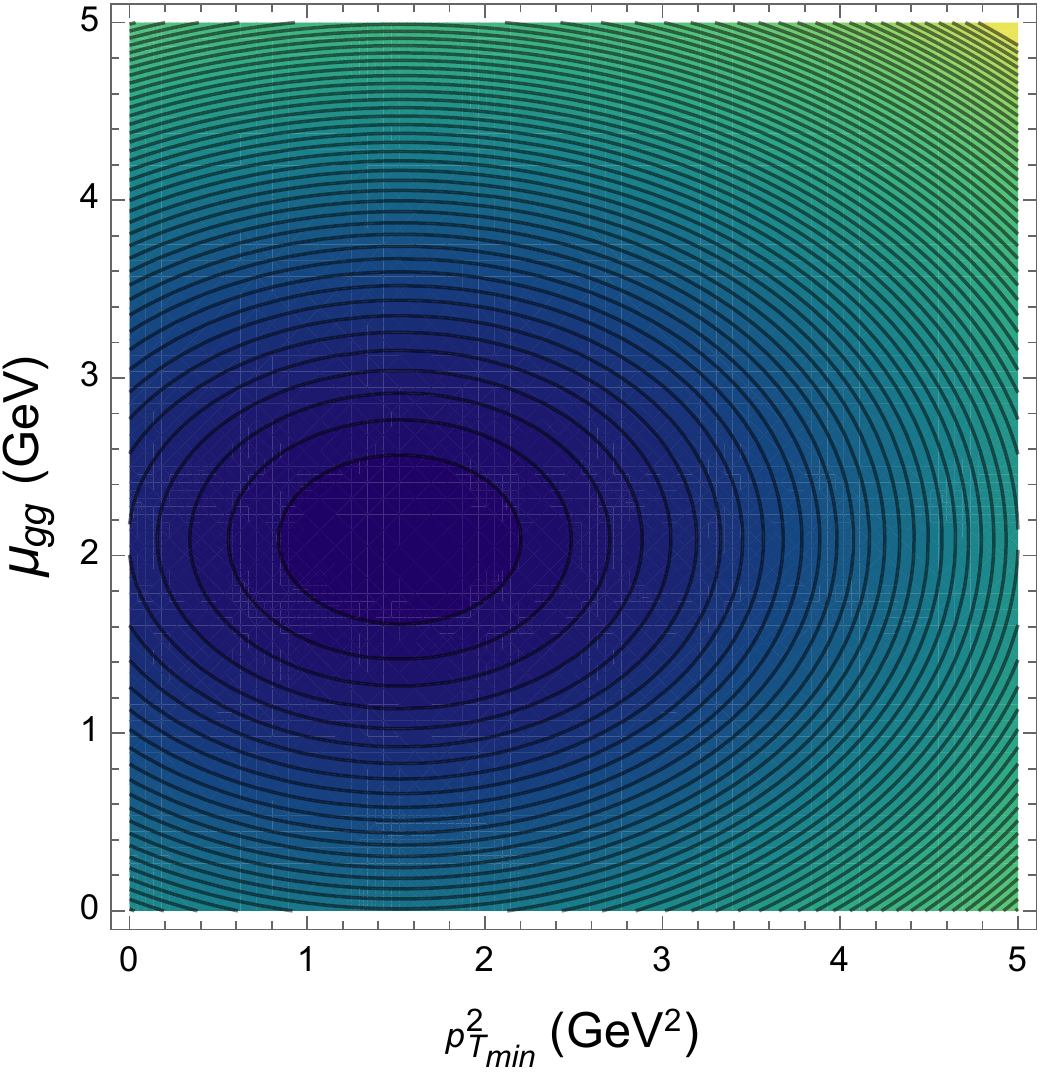}
	\caption{Contour plot of the nuclear CE as a function of the free parameters $p_{T_{min}}^{2}$ and $\upmu_{gg}$, $\upmu_{q\bar{q}} = 0.70\, {\rm GeV}$ fixed. The center
of the marine blue ellipsis-like curve corresponds to
the values of the free parameters $p_{T_{min}}^{2} = 1.531\, {\rm GeV}^{2}$ and $\upmu_{gg}=2.009\, {\rm GeV}$, for
the global minimum CE (1.531, 2.009) = 1.0031 nat.}
	\label{ff3}
\end{figure}

\begin{figure}[h]
	\centering
	\includegraphics[width=4.8cm]{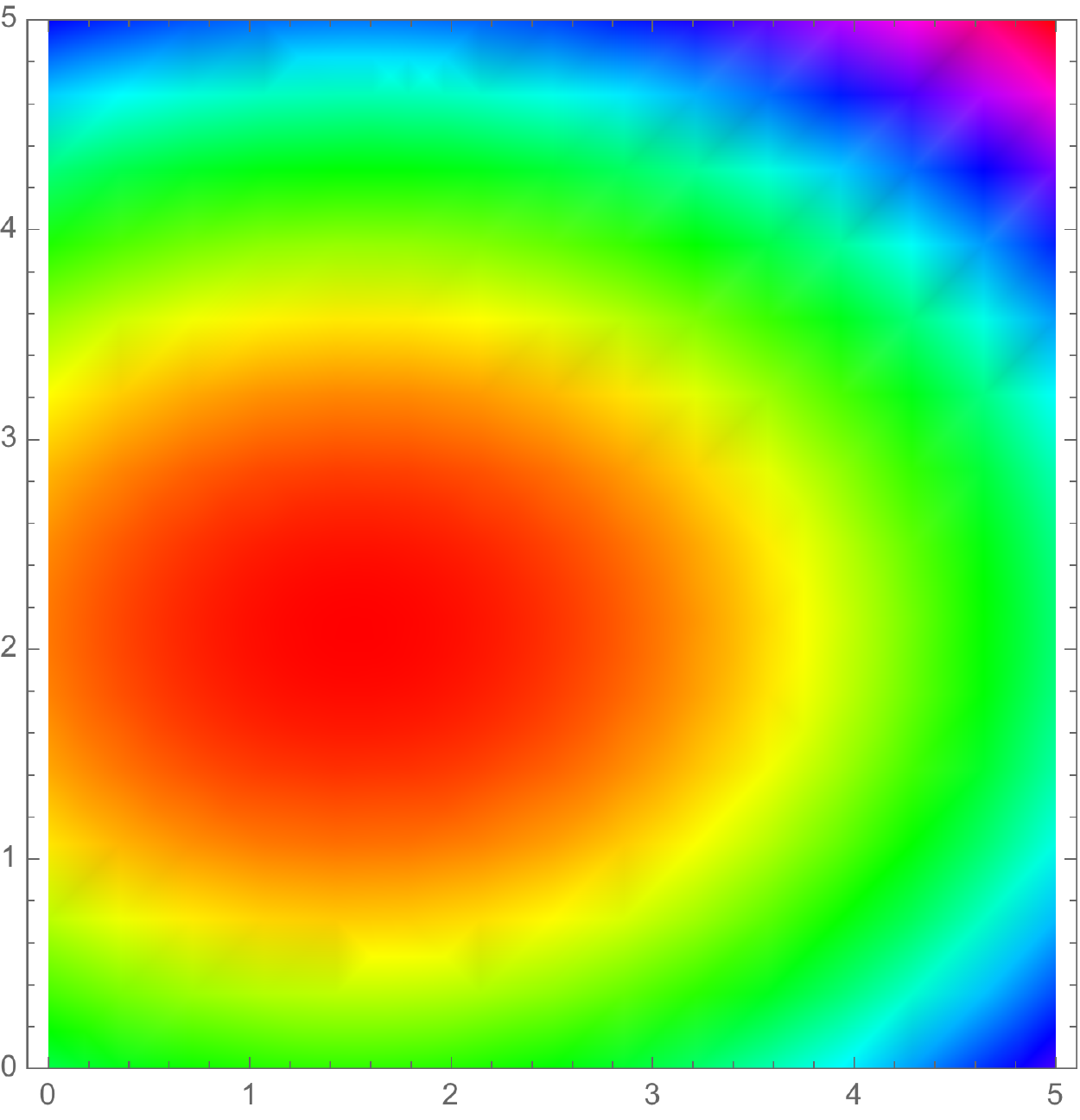}
	\caption{Density plot of the nuclear CE as a function of the free parameters $p_{T_{min}}^{2}$ ($x$ axis) and $\upmu_{gg}$ ($y$ axis), $\upmu_{q\bar{q}} = 0.70\, {\rm GeV}$ fixed.
The center
of the marine blue ellipsis-like curve corresponds to
the values of the free parameters $p_{T_{min}}^{2} = 1.531\, {\rm GeV}^{2}$ and $\upmu_{gg}=2.009\, {\rm GeV}$, for
the global minimum CE (1.531, 2.009) = 1.0031 nat.}
	\label{ff2}
\end{figure}

\begin{figure}[h]
	\centering
	\includegraphics[width=7.8cm]{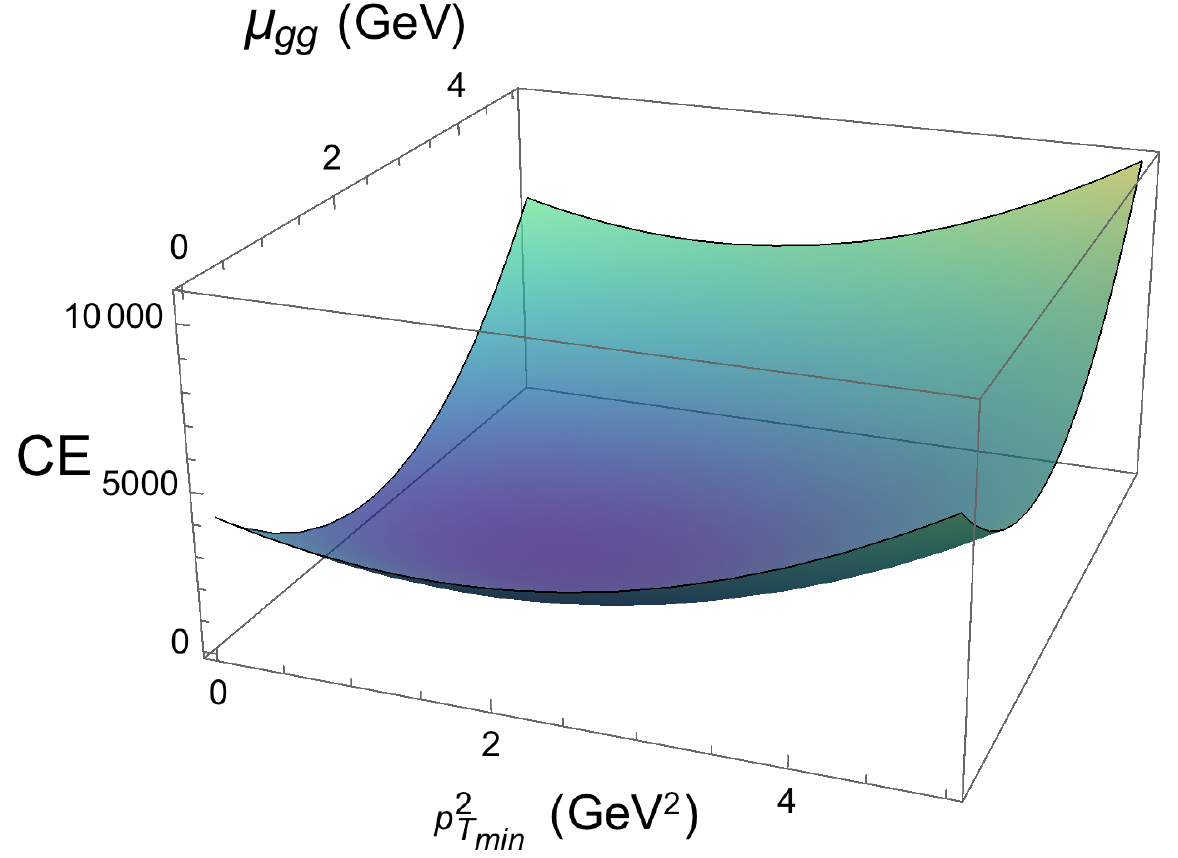}
	\caption{3D plot of the nuclear CE as a function of the free parameters $p_{T_{min}}^{2}$ ($x$ axis) and $\upmu_{gg}$ ($y$ axis), $\upmu_{q\bar{q}} = 0.70\, {\rm GeV}$ fixed.
The center
of the marine blue ellipsis-like curve corresponds to
the values of the free parameters $p_{T_{min}}^{2} = 1.531\, {\rm GeV}^{2}$ and $\upmu_{gg}=2.009\, {\rm GeV}$, for
the global minimum CE (1.531, 2.009) = 1.0031 nat.}
	\label{ff1}
\end{figure}

Thus, we calculate the modal fraction by the following equation:
\begin{equation}\label{modall}
f_{\upsigma(k)}=\frac{\vert \upsigma(k) \vert^2}{\int_{\mathbb{R}}\vert \upsigma(k)\vert ^2 dk}.
\end{equation}
The modal fraction constituting the nuclear CE specifies the probability of the reaction and thus can give us all the information we need concerning the given reaction at a given energy regime.
The mechanism of the high energy collision can be completely determined by the number of critical points of the nuclear CE to obtain the parameters, which describe any localized system \cite{Gleiser:2012tu,Gleiser:2013mga}.
After all, we apply appropriate expression for the nuclear CE \cite{Gleiser:2012tu} in order to calculate the corresponding critical points:
\begin{equation}
\label{333}
{\rm CE} =  - \int_{\mathbb{R}} f_{\upsigma(k)} \log  f_{\upsigma(k)}d k.
\end{equation}
The nuclear CE has units of nat (the natural unit of information).
Thus, having as a starting point the total inelastic cross section for hadronic and photonic reaction \cite{Giannini} given by Eqs. (\ref{siggammaptot}, \ref{siggammagammatot}), we compute the nuclear CE using Eqs. (\ref{34}) - (\ref{333}).
The results for the inelastic $pp$ and $\bar p p$ cross sections
(Eq. (\ref{sigpptot})) are shown in Figs. \ref{ff3} -- \ref{ff1}.

In Fig. \ref{ff2}, the density plot shows a predominance of 
quantum states with parameters $p_{T_{min}}^{2} = 1.531\, {\rm GeV}^{2}$ and $\upmu_{gg}=2.009\, {\rm GeV}$, in the space of parameters. Such a dominance of states fall out  in the space of parameters as one move away from the center $p_{T_{min}}^{2} = 1.531\, {\rm GeV}^{2}$ and $\upmu_{gg}=2.009\, {\rm GeV}$ that corresponds to the global minimum of the nuclear CE. In Fig. \ref{ff2}, hot [cold] colors suggest higher [lower] probability of dominant states. 

Figs. \ref{ff3} - \ref{ff1} represents the results of our calculation for the total inelastic cross section in hadron-hadron and photon-hadron or photon-photon collisions.
The numerical analysis involving Figs. \ref{ff3} - \ref{ff1} is a direct
result of the nuclear CE computation, yielding the  $p_{T_{min}}^{2}=1.531\,{\rm GeV}^{2}$ and $\upmu_{gg}=2.009\, {\rm GeV}$.

One can see from  Fig. \ref{ff3} that the concentric curves are configurationally isentropic, consisting of the states in the space of parameters that present the similar values of nuclear CE. This is accomplished in the picture of the interaction  the calculation for nuclear CE for the total inelastic cross sections where the minimum on the curve coincides with the fitted data from Ref. \cite{Giannini}. 
When compared to the experimentally obtained parameters, $p_{T_{min}}^{2} = 1.51\, {\rm GeV}^{2}$, $\upmu_{gg}=2.00\, {\rm GeV}$, we have the difference between the CE-derived values and the experimental ones given respectively by $\Delta p_{T_{min}}^{2} = 1.39\%$ and $\Delta \upmu_{gg} = 0.45\%$.
We have been able thus to determine the critical point of the nuclear CE, corresponding to the natural choice of the total inelastic cross section for the free parameters $p_{T_{min}}^{2}$ and $\upmu_{gg}$, corroborating to experimental data to a high degree of precision. 
The calculations for the nuclear CE 
derived the global minimum CE (1.531, 2.009) = 1.0031,
at the point $p_{T_{min}}^{2}=1.531\,{\rm GeV}^{2}$ and $\upmu_{gg}=2.009\, {\rm GeV}$.
At the obtained critical point, the colliding system presents a higher configurationally stability, and the calculation via the nuclear CE corroborates to it.
The main results of our calculation of high energy collision via hadron or/and photon interaction are  successfully confirmed by  the global minimum of the nuclear CE. Such a global minimum of the CE refers to the most dominant state of the nuclear configuration.
The outer regions in Fig. \ref{ff3} and Fig. \ref{ff1} show that are  no other global minima of total inelastic cross sections at the observed parameters $p_{T_{min}}^{2}$ and  $\upmu_{gg}$. 
One can also see in Figs. \ref{ff3} and \ref{ff2} the ellipsis-like closed strips which are exactly the configurational isentropic subdomains.
The center of the marine blue ellipsis-like curve corresponds to
the values of the free parameters $p_{T_{min}}^{2} = 1.531\, {\rm GeV}^{2}$ and $\upmu_{gg}=2.009\, {\rm GeV}$, for
the global minimum CE (1.531, 2.009) = 1.0031 nat.
The inner regions of the plots reflect the lower values of the nuclear CE, and regions which lie far from the center correspond to higher values of the nuclear CE.
We intend to investigate further and deeper other types of nuclear configurations, with other field theoretical effects and other wave functions, including new fermionic ones, like the ones proposed in Ref. \cite{Correa:2016pgr,Bonora:2014dfa}.

\section{Conclusions}

We present the eikonalized mini-jet model to study the energy dependence of the total hadronic and photonic cross sections.
The dependence of the cross sections has been considered through the transverse momentum of the nuclear matter distribution.
It also has been considered that the interaction of the partons obey the Poisson distribution and the average number of the inelastic collisions can be represented as a sum of soft and hard components, which depend on the center-of-mass energy and the impact parameter of the collision.
Within the eikonal limit using the Glauber multiple collision approximation, the total inelastic $pp$ and $p\bar p$ scattering and $\gamma{p}$ and $\gamma\gamma$ cross sections were estimated
from the corresponding inelastic nucleon-nucleon cross section.
During the calculation procedure, the global minima of the configurational entropy have been detected. Such minima show the critical point which can define within the uncertainty of the calculation the total inelastic production cross section for hadrons and/or $\gamma$ induced reactions.
Within the Color - Glass Condensate mini-jet model it has been possible to provide the natural choice used on the complex theoretical analysis.
Our calculations predicted the values of the free fitted parameters
$p_{T_{min}}^{2} = 1.531\, {\rm GeV}^{2}$ and $\upmu_{gg}=2.009\, {\rm GeV}$, for
the global minimum CE (1.531, 2.009) = 1.0031 nat, within the uncertainties of calculation, $\Delta p_{T_{min}}^{2} = 0.7\%$, whereas $\Delta \upmu_{gg} = 0.45\%$.
The errors represent its upper limits for the given fitted parameters for the given center-of-mass energy and impact parameter value.
The deep investigation of the critical point of CE the interacted system gives us information about the stability of the system and the number of nuclear states available. Such a result can be successfully used to obtain the set of the trustable parameters which can adequately describe interaction as a whole.

\acknowledgments GK thanks to The S\~ao Paulo Research Foundation -- FAPESP (grant No. 2018/19943-6).

\end{document}